\documentclass[aps,pre,amssymb,twocolumn,superscriptaddress]{revtex4-1}
\usepackage{graphicx}
\usepackage{amsmath}

\begin{document}
\title{Free utility model for explaining the social gravity law}
\author{Hao Wang}
\affiliation{Institute of Transportation System Science and Engineering, Beijing Jiaotong University, Beijing {\rm 100044}, China}

\author{Xiao-Yong Yan}
\email{yanxy@bjtu.edu.cn}
\affiliation{Institute of Transportation System Science and Engineering, Beijing Jiaotong University, Beijing {\rm 100044}, China}
\affiliation{Comple$\chi$ Lab, University of Electronic Science and Technology of China, Chengdu {\rm 611731}, China}

\author{Jinshan Wu}
\affiliation{School of Systems Science, Beijing Normal University, Beijing {\rm 100875}, China}

\begin{abstract}
Social gravity law widely exists in human travel, population migration, commodity trade, information communication, scientific collaboration and so on. Why is there such a simple law in many complex social systems is an interesting question. Although scientists from fields of statistical physics, complex systems, economics and transportation science have explained the social gravity law, a theoretical explanation including two dominant mechanisms, namely individual interaction and bounded rationality, is still lacking. Here we present a free utility model, whose objective function is mathematically consistent with the Helmholtz free energy in physics, from the perspective of individual choice behavior to explain the social gravity law. The basic assumption is that bounded rational individuals interacting with each other will trade off the expected utility and information-processing cost to maximize their own utility. The previous explanations of the social gravity law including the maximum entropy model, the free cost model, the Logit model and the destination choice game model are all special cases under our model. Further, we extend the free utility model to the network. This model not only helps us to better understand the underlying mechanisms of spatial interaction patterns in complex social systems, but also provides a new perspective for understanding the potential function in game theory and the user equilibrium model in transportation science.
\end{abstract}

\maketitle

\section{Introduction}\label{sec:1}
Predicting the mobility of people, goods and information between locations is an important problem in fields as diverse as sociology \cite{Garra16}, economics \cite{Niedercorn69}, demography \cite{Roy03}, epidemiology \cite{Jia20}, transportation science \cite{Dios11} and network science \cite{Bar11}. 
For more than a hundred years, scholars have proposed a variety of models to predict such mobility \cite{Dios11,Bar11,Stouffer40,Sen12,Simini12,Yan14,Yan17,Liu20}. These models are named spatial interaction models in economics \cite{Sen12} and trip distribution models in transportation science \cite{Dios11}. The gravity model is the most influential mobility prediction model and has been applied in many fields \cite{Bar11}. For example, it is used to predict population migration \cite{Fagiolo13}, commodity trade \cite{Karpiarz14}, commuting flows \cite{Masucci13} and public transportation flows \cite{Masucci13,Goh12}. The gravity model is popular because spatial interaction phenomena in these fields all obey a law known as the social gravity law under which the flow between two locations is proportional to the activity (usually quantified by population, GDP and so on) of these two locations and decays with the power function of the distance between them, similar to Newton's law of universal gravitation.
As early as 1846, Desart found that the railway passenger flow between stations in Belgium obeyed the social gravity law \cite{Odlyzko15}. This may be the earliest record of the discovery of the social gravity law. Later, Carey \cite{Carey58} and Ravenstein \cite{Ravenstein89} found that population migrations of the United States and the United Kingdom, respectively, obeyed the social gravity law. Additionally, Reilly \cite{Reilly29} found that retail business drawn from cities obeyed the social gravity law. 
In recent years, with the continuous development of modern electronic and information technology, there have been many approaches (such as GPS, mobile phones and social networking sites) to recording the mobility data of people, goods and information over long periods. By analyzing these data, scientists have found many phenomena showing obedience to the social gravity law in various systems  \cite{Bar11}. 
For example, commuting flow in the United States \cite{Viboud06}, highway traffic flow in Korea  \cite{Jung08},  telecommunication flow in Belgium \cite{Krings09}, global airline flow \cite{Balcan09}, global cargo ship movement \cite{Kaluza10} and even global scientific collaboration \cite{Pan12} followed the social gravity law.

Why does such a simple law apply to so many complex social systems? This problem has long excited the curiosity of many scholars. In the past half century, scientists have proposed different explanations of the roots of the social gravity law, among which Wilson's maximum entropy model \cite{Wilson67} is the prevailing explanation. He proposed an approach to deriving the doubly-constrained gravity model by maximizing the entropy of the trip distribution between locations in the transportation system under the constraints of the total cost, the trip production volume and attraction volume of each location. 
However, unlike the total amount of internal energy available to all gas molecules in a system, which is determined exogenously, the total cost cannot be estimated in a real transportation system \cite{Hua79}. 
Tomlin and Tomlin \cite{Tomlin68} constructed a free cost model by analogy with the Helmholtz free energy in physics. This model can lead to the same results as Wilson's maximum entropy model but does not need the prior constraint of the total cost.
However, both the maximum entropy model and the free cost model are macroscopic explanations of the social gravity law. They provide only the most probable macroscopic distribution state but do not take into account individual choice behavior in destination choice \cite{Sheppard78}.
On the other hand, economists have explained the social gravity law by describing individual choice behavior using utility theory, in which the most influential study is that of Domencich and McFadden, who applied random utility theory to model individual destination choice behavior \cite{DoMc75}. They assumed that the traveler always selects the destination with the highest utility, but her perception of the destination utility exhibits random error. If these errors follow the independent and identical Gumbel distribution, the Logit model \cite{Hensher18} can be derived. If the destination utility consists only of the destination attractiveness and the travel cost to the destination, the Logit model can derive the singly-constrained gravity model.
However, all the aforementioned models neglect individual interaction, which is a ubiquitous phenomenon in many complex systems \cite{Vicsek12}. For example, in a real transportation system, a traveler considers not only the constant values of destination attractiveness and travel cost but also possible crowding at the destination and congestion on the way. 
Recently, Yan and Zhou \cite{Yan19} modeled the individual destination choice process as a congestion game, including interaction among travelers, and further derived the singly-constrained gravity model. However, they assumed that all travelers are perfectly rational and can accurately perceive the utilities of all destinations. In practice, individual rationality is bounded because of the intractability of the alternative choice problem and limited information-processing resources \cite{Simon72}. However, an explanation of the social gravity law that simultaneously reflects individual interaction and bounded rationality is still lacking.

In this paper, we develop a free utility model from the perspective of individual choice behavior to explain the social gravity law in the context of the destination choice problem in transportation science. First, we establish the individual choice model on the basis of how a bounded rational traveler trades off expected utility and information-processing cost. Then, we extend the individual choice model to the collective choice model, including infinite noninteractive or interactive travelers, and further derive the gravity model from the collective model. We next extend the collective model of interactive travelers, named the free utility model, to a network. Finally, we contrast the similarities and differences between the free utility model and the free energy in physics and further discuss the potential application value of the free utility model in different scientific fields. 

\section{Model}\label{sec:2}
{
\subsection{Gravity model}
Before presenting our model to explain the social gravity law, we first briefly introduce the gravity model that is widely used to predict the spatial interaction flow obeying the social gravity law.
The earliest gravity model \cite{Odlyzko15}, generated from an analogy with Newton's law of universal gravitation, has the following functional form:
\begin{equation} \label{eq0-1}
T_{ij} = \alpha \frac{m_i m_j}{d_{ij}^\beta},
\end{equation}
where $T_{ij}$ is the flow from origin $i$ to destination $j$, $m_i$ is the activity (usually quantified by population) of location $i$, $d_{ij}$ is the distance between $i$ and $j$, and $\alpha$ and $\beta$ are parameters.
This original gravity model has a simple form and can be used to reproduce the distribution pattern of spatial interaction flow, but difficulties arise when predicting future flows. 
For example,  the $T_{ij}$ calculated by Eq. (\ref{eq0-1}) would increase by a factor of 100 if the future populations $m_i$ and $m_j$ both increase by factors of 10, which is clearly not realistic.

An improvement to overcome this shortcoming of the original gravity model is to discard the fixed parameter $\alpha$ and use the following equation to calculate the flow:
\begin{equation} \label{eq0-2}
T_{ij}  = O_i  \frac{ m_j d_{ij}^{-\beta} }{\sum_j m_j d_{ij}^{-\beta}},
\end{equation}
where $O_i$ is the outflow of location $i$, which is usually roughly proportional to the population of location $i$, i.e., $O_i \approx \theta m_i$  \cite{Simini12}.
The results calculated by Eq. (\ref{eq0-2}) satisfy the constraint $\sum_j T_{ij}=O_i$, so Eq. (\ref{eq0-2})   is named
the singly-constrained gravity model  \cite{Dios11}.

The more commonly used gravity model in transportation science is the doubly-constrained gravity model \cite{Dios11}, which is constructed to satisfy two constraints, i.e., $\sum_j T_{ij}=O_i$ and $\sum_i T_{ij}=D_j$, where $D_j$ is the inflow of location $j$. 
In addition, travel cost $c_{ij}$ is used more often than distance $d_{ij}$ in trip distribution  forecasting, so the doubly-constrained gravity model can be written as
\begin{equation} \label{eq0-3}
T_{ij}  =a_i b_j O_i  D_j f(c_{ij}),
\end{equation}
where $f(c_{ij})$ is the travel cost function and $a_i  =  1/{\sum_j b_j D_j f(c_{ij})}$ and $b_j  =  1/{\sum_i a_i O_i f(c_{ij})}$ are balancing factors.

Next, we will establish the free utility model from the perspective of individual choice behavior  to derive the singly- and doubly-constrained gravity models and explain the social gravity law.
}

\subsection{Individual choice model}

Similar to some of the aforementioned studies explaining the social gravity law, we conduct our research in the context of individual destination choice behavior in the transportation system.
In this system, there are $M$ origins labeled $i$ $(i=1, 2, \dots , M)$ and $N$ destinations labeled $j$ $(j=1, 2, \dots , N)$. $O_{i}$ is the number of trips leaving origin $i$. 
We initially start with the simplest system, in which there is only one origin $(M=1)$ and one trip $(O_{i}=1)$ made by a traveler who can select $N$ $(N>1)$ destinations. The utility of destination $j$ for this traveler is $u_{ij}$, which describes her satisfaction degree concerning the attractiveness of destination $j$ and the travel cost from origin $i$ to destination $j$.
According to the utility maximization principle, this traveler will select the destination with the highest utility only if she is perfectly rational \cite{Fishburn82}. However, it is difficult for a bounded rational traveler to exactly perceive the utilities of all destinations in reality. In this case, her choice is probabilistic, i.e., destination $j$ is selected with probability $p_{ij}$, which depends on $u_{ij}$ \cite{Luce59}. 
If the traveler is completely uncertain about the utilities of all destinations, she can select destination $j$ only with probability $p_{ij}=\frac{1}{N}$, and her expected utility $\sum_j p_{ij}u_{ij}$  is the average of the utilities of all destinations. 
If she wants to obtain higher expected utility, she must acquire knowledge about the utilities of the destinations through information processing \cite{Marsili99}. A natural measure of information processing is negative information entropy $-H=\sum_j p_{ij}\ln p_{ij}$ \cite{Marsili99,Wolpert12,Ortega13}. Assuming that the price of unit information is $\tau$, then the information-processing cost is $-\tau H$. 
If $\tau \to \infty$, which means that the information-processing cost is extremely high, the traveler does not care about the expected utility and focuses only on the information-processing cost. Hence, she will make a totally random choice of destination, that is, a uniform distribution over the set of destinations. 
If $\tau=0$, which means that information processing has no cost, she will select only the destination with the highest utility. 
In general case $\tau>0$, she must trade off her expected utility and the information-processing cost \cite{Guan20, Tkacik16} to achieve the total utility maximization goal, i.e.,
\begin{equation}
\label{eq1}
\begin{aligned}
\max w=& \sum_j p_{ij}u_{ij}+\tau H, \\
	\mathrm{s.t.} \quad &\sum_j p_{ij}=1,
\end{aligned}
\end{equation}
where $\sum_j p_{ij}u_{ij}+\tau H$ is the objective function subjected to $\sum_j p_{ij}=1$. Using the Lagrange multiplier method, we can obtain the solution of Eq.~(\ref{eq1}) as follows:
\begin{equation}
\label{eq2}
L(p_{ij},\lambda)=\sum_j  p_{ij}u_{ij}+\tau H-\lambda(\sum_j p_{ij}-1),
\end{equation}
where $\lambda$ is a Lagrange multiplier. According to $\frac{\partial L}{\partial p_{ij}}=0$ for all destinations, we can obtain
\begin{equation}
\label{eq3}
u_{ij}-\tau(\ln p_{ij}+1)=\lambda,
\end{equation}
which means that all destinations have the same utility minus the marginal information-processing cost under the traveler's optimal choice strategy. This situation is very similar to consumer equilibrium in microeconomics \cite{Tewari03}, in which the marginal utility of each good is equal. Thus, we name $u_{ij}-\tau(\ln p_{ij}+1)$  the \emph{marginal utility} of destination $j$. The total utility of the system is the sum of the integrals of the marginal utilities of all destinations. The optimal choice strategy for the traveler is to follow the \emph{equimarginal principle} \cite{Tewari03} to select destinations in order to achieve maximum total utility. Combining Eq.~(\ref{eq3}) and $\sum_j p_{ij}=1$, we can derive
\begin{equation}
\label{eq4}
p_{ij}=\frac{\mathrm{e}^{u_{ij} / \tau}}{\sum_j \mathrm{e}^{u_{ij} / \tau}},
\end{equation}
which is the equilibrium solution of Eq.~(\ref{eq1}). 
{ In economics, the reciprocal of the parameter $\tau$ in Eq. (\ref{eq4}) is commonly referred to as the intensity of choice parameter \cite{Brock97}, which measures how sensitive a traveler is with respect to differences in the destination utilities. The greater $1/\tau$ is, the more sensitive the traveler is to differences in the destination utilities, and vice versa.}
In mathematical form, Eq.~(\ref{eq4}) is the Logit model derived in terms of random utility theory \cite{DoMc75}. However, our derivation does not need to assume in advance that the destination utility perception errors follow the Gumbel distribution. 

\subsection{Collective choice model}

We further extend the transportation system to the case where origin $i$ has infinite homogeneous travelers (i.e., $O_i\gg1$), each of whom still has only one trip. The question then becomes how these trips are distributed among various destinations.
If the utility of destination $j$ is not affected by the number of trips to $j$, Eq.~(\ref{eq4}) can still be applied in this system. On this occasion, the number of trips from $i$ to $j$ is $T_{ij}=O_{i}p_{ij}$, and the system entropy $S$ is the sum of the information entropy $H$ of each trip{, i.e., $S=O_i H$}. Therefore, we can rewrite Eq.~(\ref{eq1}) as
\begin{equation}
\label{eq5}
\begin{aligned}
\max W=& \sum_j T_{ij}u_{ij}+\tau S, \\
	\mathrm{s.t.} \quad &\sum_j T_{ij}=O_{i}.
    \end{aligned}
\end{equation}
Similarly, all homogeneous travelers follow the equimarginal principle so that all destinations have the same marginal utility for them.  If the utility of destination $j$ is abstractly written as $u_{ij}=A_{j}-c_{ij}$, where $A_j$ is the constant attractiveness of destination $j$ reflecting the activity opportunities (including variables such as retail activity and employment density) \cite{Sheffi85} available there, and $c_{ij}$ is the constant travel cost from $i$ to $j$,  we can use the Lagrange multiplier method to obtain the equilibrium solution of Eq.~(\ref{eq5}) as
\begin{equation}
\label{eq6}
T_{ij}=O_{i}\frac{\mathrm{e}^{(A_j-c_{ij}) / \tau}}{\sum_j \mathrm{e}^{(A_j-c_{ij}) / \tau}},
\end{equation}
which is the same as the singly-constrained gravity model derived from the Logit model \cite{DoMc75}.

However, in practice, the utility of destination $j$ is affected by the number of trips from $i$ to $j$ \cite{Fujita89}; that is, $u_{ij}$ is a function of $T_{ij}$. For example, { an increase in }the trip number $T_{ij}$ from origin $i$ to destination $j$ will {increase} the travel cost from $i$ to $j$ and { decrease} the attractiveness of destination $j$ {\cite{Yan19}}, both of which will change the utility of destination $j$. Hence, the utility can be abstractly written as $u_{ij}(T_{ij})=A_j-l_j(T_{ij})-c_{ij}-g_{ij}(T_{ij})$, where $l_j(T_{ij})$ is the variable attractiveness function of destination $j$ and $g_{ij}(T_{ij})$ is the congestion function on the way from $i$ to $j$.  { To solve the subsequent optimization problems, we assume that  $l_j(T_{ij})$ and $g_{ij}(T_{ij})$ are both monotonically nondecreasing differentiable functions.}
We already know that each traveler follows the equimarginal principle to make the optimal choice. In this context, a traveler's choice of destination depends on how other travelers are distributed over all destinations \cite{Conlisk76}. The phenomenon of one individual's behavior being dependent on the behavior of other individuals, known as individual interaction, is widespread in social systems. In this interactive system, the optimal choice strategy for all travelers is still to assign all destinations the same marginal utility. 
Since the total utility of the system is the sum of the integrals of the marginal utilities of all destinations, the utility maximization model of the interactive system can be written as
\begin{equation}
\label{eq7}
\begin{aligned}
\max W=& \sum_j \int_0^{T_{ij}} u_{ij}(x) \mathrm{d} x+\tau S, \\
	\mathrm{s.t.} \quad &\sum_j T_{ij}=O_{i}.
    \end{aligned}
\end{equation}
Interestingly, the objective function of Eq.~(\ref{eq7}) is mathematically consistent with the Helmholtz free energy in physics \cite{Kittel80}; thus, we name Eq.~(\ref{eq7}) the \emph{free utility} model. Analogous to the thermodynamic system, the equilibrium trip distribution maximizes the free utility function in the transportation system.
In addition, the first term of the free utility function, $\sum_j \int_0^{T_{ij}} u_{ij}(x) \mathrm{d} x$, is analogous to internal energy, including potential energy in a thermodynamic system. Monderer and Shapley \cite{Monderer96} therefore named this term the \emph{potential function}. The free utility model with $\tau=0$ is essentially a potential game with an infinite number of agents. 

\begin{figure}
	\centering
	\includegraphics[width=1.0\columnwidth]{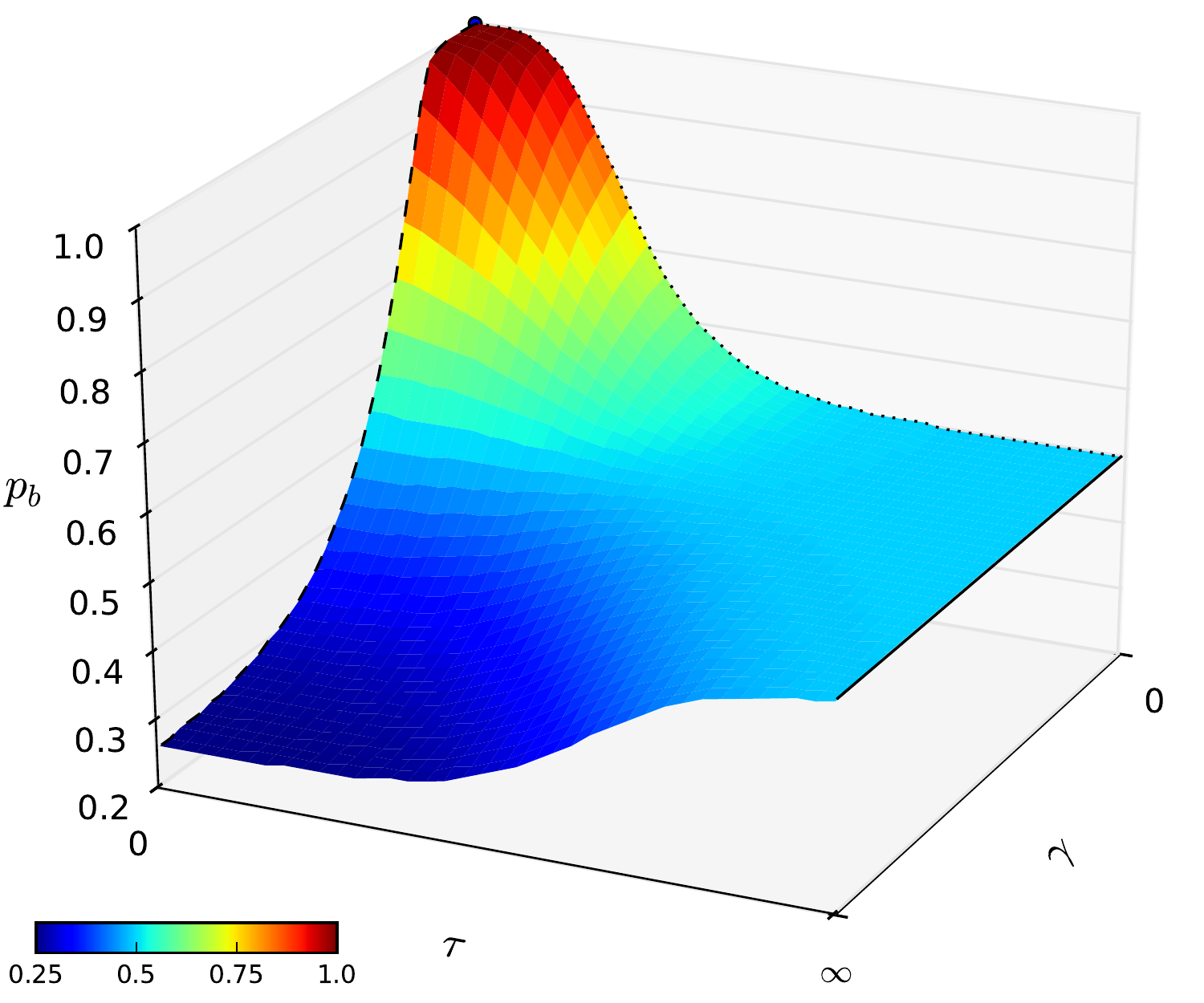}
	\caption{{\bf The effect of parameter changes on the optimal solution of the free utility model.} In this simple system, there is only one origin, labeled \emph{a}, and two destinations, labeled \emph{b} and \emph{c}. The origin has 200 travelers, each of whom has only one trip. The utilities of  \emph{b} and \emph{c} are $u_{ab} = 20-3\gamma T_{ab}$ and $u_{ac} = 2-\gamma T_{ac}$, respectively, where $T_{ab}$ and $T_{ac}$ are the number of travelers selecting destinations \emph{b} and \emph{c}, respectively, and $\gamma$ is the interaction strength parameter. The information-processing cost for these travelers is $-\tau S$, where $\tau$ is the information-processing price. The surface of the figure describes probability $p_b$ that destination \emph{b} is selected in the optimal solution of the free utility model under different parameter combinations. }
	\label{fig1}
\end{figure}

Here, to generate gravity-like behavior, the utility needs a logarithmic dependence on the number of travelers from $i$ to $j$, i.e., $u_{ij}(T_{ij})=A_{j}-c_{ij}-\gamma \ln T_{ij}$, where $\gamma \ln T_{ij}=l_{j}(T_{ij})+g_{ij}(T_{ij})$ is the simplified cost function, including the travel congestion cost and the destination variable attractiveness, and $\gamma$ is a non-negative parameter. 
{The logarithmic form of the cost function can be explained by the Weber-Fechner law  in behavioral psychology \cite{Takemura14}. This law holds that the magnitude of human perception (e.g., the traveler's perception of travel cost) is proportional to the logarithm of the magnitude of the physical stimulus (e.g., the number of travelers $T_{ij}$).}
Equation~(\ref{eq7}) can be specifically written as
\begin{equation}
\label{eq8}
\begin{aligned}
\max W=& \sum_j \int_0^{T_{ij}} (A_{j}-c_{ij}-\gamma \ln x) \mathrm{d} x+\tau S, \\
	\mathrm{s.t.} \quad &\sum_j T_{ij}=O_{i}.
    \end{aligned}
\end{equation}
Using the Lagrange multiplier method, we can obtain
\begin{equation}
\label{eq9}
T_{ij}=O_{i}\frac{\mathrm{e}^{(A_j-c_{ij}) / (\gamma+\tau)}}{\sum_j \mathrm{e}^{(A_j-c_{ij}) / (\gamma+\tau)}}.
\end{equation}
This result is the singly-constrained gravity model \cite{Dios11}. Parameter $\tau$ reflects the information-processing price for bounded rational travelers, and parameter $\gamma$ reflects travelers' interaction strength. Figure~\ref{fig1} shows the effect of changes in parameters $\tau$ and $\gamma$ on the free utility optimal solution in a simple system with one origin and two destinations. The surface in Fig.~\ref{fig1} describes the optimal solution of the free utility model in this simple system. 
From Eq.~(\ref{eq8}), we can see that when $\gamma > 0$ and $\tau = 0$ (meaning no information-processing cost), the free utility model is equivalent to the degenerated destination choice game model \cite{Yan19} in terms of potential game theory, as shown in the left dashed line of Fig.~\ref{fig1}; 
when $\gamma >0$ and $\tau \to \infty$ (meaning that the information-processing cost is too high), travelers can only uniformly randomly select each destination, as shown in the lower-right solid line of Fig.~\ref{fig1};
when $\tau > 0$ and $\gamma = 0$ (meaning that there is no interaction among travelers), the free utility model is equivalent to the Logit model \cite{DoMc75}, as shown in the upper-right dotted line of Fig.~\ref{fig1}; and
when $\tau = 0$ and $\gamma = 0$, all travelers will select the destination with the highest constant utility, as shown in the intersection point of the dashed line and dotted line of Fig.~\ref{fig1}.

Above, we considered only a simple system with a single origin. However, there is more than one origin $(M > 1)$ in a real transportation system. In such a system, the utility form of destination $j$ is changed. It is affected by not only the number of trips from $i$ to $j$ but also the total number of trips attracted to destination $j$. Hence, the utility of destination $j$ for travelers at origin $i$ can be abstractly written as  $u_{ij}(T_{ij}, D_j)=A_j-l_j(D_j)-c_{ij}-g_{ij}(T_{ij})$, where $D_j$ is the total number of trips attracted to destination $j$, i.e., $D_j=\sum_i T_{ij} ${, and used as the independent variable  of the variable attractiveness function $l_j(D_j)$  of destination $j$}. If $D_j$ is as constant as $O_i$, that is, not only the total number of trips emanating from origin $i$ but also the number of trips {attracted to} any destination is fixed,
then, the variable attractiveness function $l_j(D_j)$ of destination $j$ is also constant and thus can be merged into the constant attractiveness $A_j$ of destination $j$. In other words, only the travel congestion function $g_{ij}(T_{ij})$ influences the destination choice behavior of travelers. If $g_{ij}(T_{ij})=\gamma \ln T_{ij}$, the free utility model of the system can be written as
\begin{equation}
\label{eq10}
\begin{aligned}
\max W=&  \sum_i \sum_j \int_0^{T_{ij}} (A_{j}-c_{ij}-\gamma \ln x) \mathrm{d} x+\tau  \sum_i S_i, \\
	\mathrm{s.t.} \quad &\sum_j T_{ij}=O_{i},\\
	&\sum_i T_{ij}=D_{j},
    \end{aligned}
\end{equation}
where the two constraints $\sum_j T_{ij}=O_{i}$ and $\sum_i T_{ij}=D_{j}$ are the fixed total number of trips emanating from origin $i$ and the fixed total number of trips attracted to destination $j$, respectively. Using the Lagrange multiplier method {(see Appendix)}, we can obtain
\begin{equation}
\label{eq11}
T_{ij}=a_{i}b_{j}O_{i}D_{j}\mathrm{e}^{-c_{ij} / (\gamma+\tau)},
\end{equation}
where $a_{i}=1/\sum_j b_j D_j \mathrm{e}^{-c_{ij} / (\gamma+\tau)}$ and $b_{j}=1/\sum_i a_i O_i \mathrm{e}^{-c_{ij} / (\gamma+\tau)}$. This is the doubly-constrained gravity model widely used in transportation science \cite{Dios11}.
This free utility model can be reduced to some classical models under specific parameter combinations. 
When $\tau > 0$ and $\gamma=0$ (meaning that there is no interaction among travelers), Eq.~(\ref{eq10}) is equivalent to the free cost model proposed by Tomlin et al. \cite{Tomlin68}, and its solution has the same form as that of Wilson's maximum entropy model \cite{Wilson67}.
When $\tau = 0$ and $\gamma = 0$, Eq.~(\ref{eq10}) is equivalent to the Hitchcock-Koopmans problem \cite{Hitchcock41}, which asks, for every origin $i$ and destination $j$, how many travelers must journey from $i$ to $j$ in order to minimize the total cost  $\sum_i \sum_j (c_{ij}-A_{j})T_{ij}$. 
When $\tau \to \infty$ (meaning that the first term of the free utility function in Eq.~(\ref{eq10}) is negligible), Eq.~(\ref{eq10}) is equivalent to the equal priori probability Sasaki model \cite{Tomlin68} with the solution $T_{ij}\propto O_iD_j$.

In a real transportation system, the travel cost $c_{ij}$ often follows an approximate logarithmic relationship with distance $d_{ij}$, i.e., $c_{ij}\approx \beta\ln d_{ij}$ \cite{Yan13}. Using $\beta\ln d_{ij}$ instead of $c_{ij}$ in Eq.~(\ref{eq11}), we can obtain the social gravity law
\begin{equation}
\label{eq12}
T_{ij}=a_{i}b_{j}O_{i}D_{j}d_{ij}^{-\beta / (\gamma+\tau)}.
\end{equation}
Thus far, we have achieved the goal of explaining the root of the social gravity law. The social gravity law is a macroscopic phenomenon caused by the interaction of bounded rational individuals in destination choice. In contrast, the destination choice game model considers only individual interaction but ignores individual bounded rationality, which is almost universal in practice, while the maximum entropy model, the free cost model and the Logit model reflect the randomness of individual choice decisions  but do not reflect individual interaction.

\subsection{Network expansion model}

\begin{figure}
	\centering
	\includegraphics[width=1.05\columnwidth]{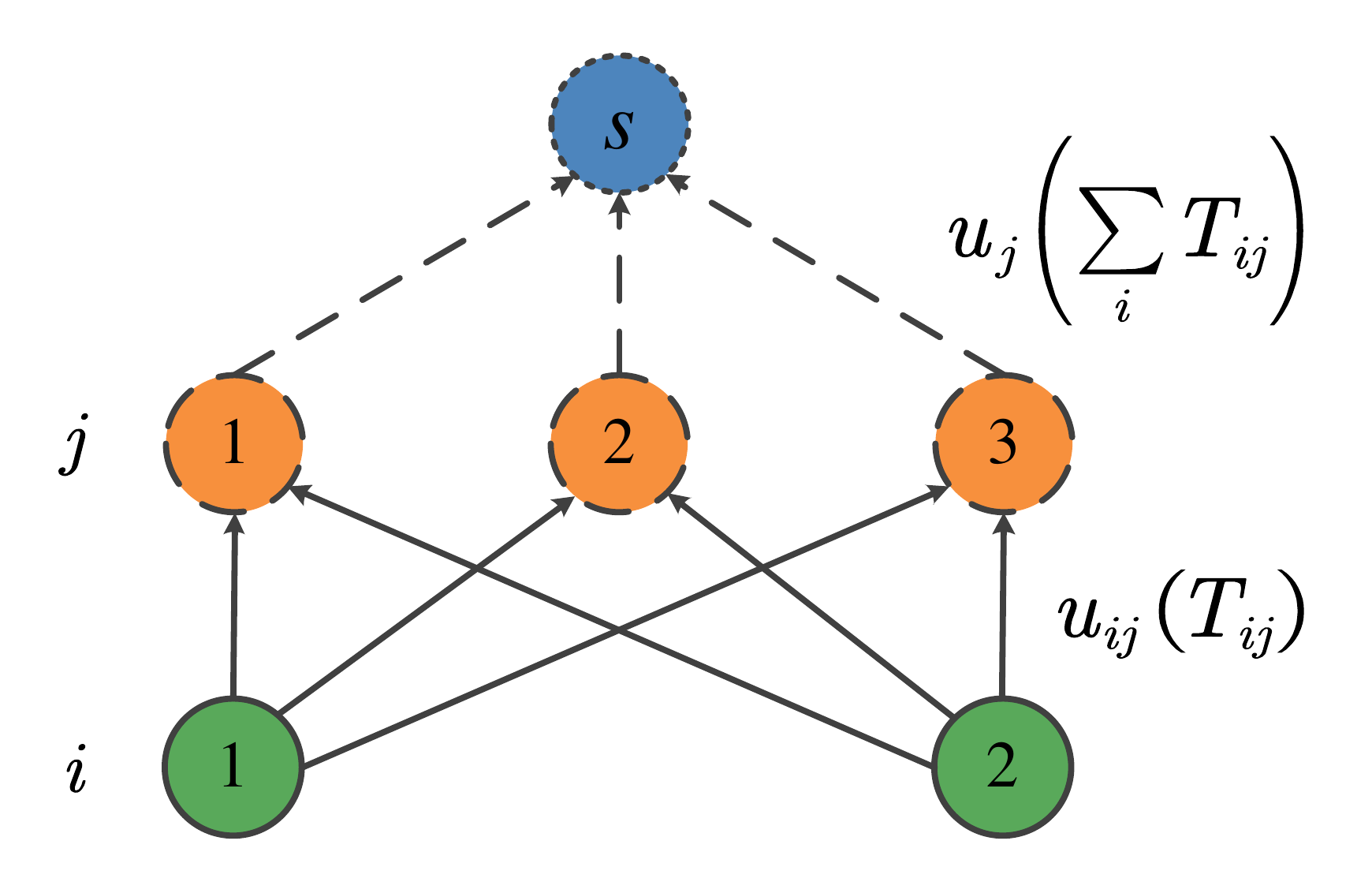}
	\caption{{\bf A{n example} network to illustrate destination choice behavior in a transportation system with multiple origin-destination pairs.} In this network, the green nodes in the bottom layer is are origin nodes, labeled $i$, and the orange nodes in the middle layer are destination nodes, labeled $j$. The solid lines from the green nodes to the orange nodes are links $ij$, with flow $T_{ij}$ and utility $u_{ij}(T_{ij})$. The blue node in the top layer is the dummy node $s$. The dashed lines from the orange nodes to the dummy node are  links $js$, with flow $\sum_i T_{ij}$ and utility $u_j(\sum_i T_{ij})$. Individual destination choice behavior is route choice behavior in this network. The path from $i$ to $s$ is composed of link $ij$ and corresponding link $js$.}
	\label{fig2}
\end{figure}

The above constraints $\sum_j T_{ij}=O_{i}$ and $\sum_i T_{ij}=D_{j}$ are essential in the classic four-step travel demand forecasting models in transportation science since the second-step model predicting trip distribution requires the input values of trip production and attraction resulting from the first-step model \cite{Dios11}.
However, the total number of trips $D_j$ attracted to destination $j$ is impossible to fix beforehand in a real transportation system since $D_j= \sum_i T_{ij}$ is the product of the individual destination choice process. In other words, $D_j$ is variable. In this case, the free utility model can still describe traveler destination choice behavior.
For a better understanding, we transform the destination choice problem in the transportation system with multiple origin-destination pairs into an equivalent route choice problem in a network, as shown in Fig.~\ref{fig2}.
In this network, there are $M$ origin nodes, $N$ destination nodes and one dummy node $s$. There are $M\times N$  links leading from each origin node to each destination node and $N$  links leading from each destination node to the dummy node. Each origin node has $O_i$ travelers whose final destination in this network is the dummy node $s$.
The traveler journeying from $i$ to $s$ has $N$ alternative paths. Each path is composed of two links, $ij$ and $js$. The flow along link $ij$ is $T_{ij}$, and that along $js$ is $\sum_i T_{ij}$. The path utility consists of the utility $u_{ij}(T_{ij})=-c_{ij}-g_{ij}(T_{ij})$ of link $ij$ and the utility $u_j(\sum_i T_{ij})=A_j-l_j(\sum_i T_{ij})$ of link $js$. Hence, the free utility model for this  network can be written as
\begin{equation}
\label{eq13}
\begin{aligned}
\max W=&  \sum_i \sum_j \int_0^{T_{ij}} u_{ij}(x) \mathrm{d} x\\
&+\sum_j \int_{0}^{\sum_i T_{ij}} u_{j}(x) \mathrm{d} x
+\tau  \sum_i S_i, \\
\mathrm{s.t.} \quad &\sum_j T_{ij}=O_{i},
\end{aligned}
\end{equation}
where $S_i=-\sum_j T_{ij}\ln \frac{T_{ij}}{O_i}$. In addition, some paths may not be selected since the number of trips $O_i$ is finite in practice; thus, constraint $T_{ij}\geq 0$ must be added to Eq.~(\ref{eq13}). 
Interestingly, Eq.~(\ref{eq13}) is identical to the stochastic user equilibrium (SUE) model \cite{Fisk80} for traffic assignment in transportation science. There are many algorithms that can solve this model and calculate the traffic flow on all links in the network \cite{Sheffi85}. Although the free utility model in Eq.~(\ref{eq13}) cannot provide an explicit expression of $T_{ij}$, it can better characterize collective destination choice behavior than the doubly-constrained gravity model that specifies the number of trips $D_j$ attracted to destination $j$.

\section{Discussion and conclusions}\label{sec:3}
In this paper, we developed a free utility model that can explain the social gravity law from the perspective of individual choice behavior. The model makes two basic assumptions: (1) the individual pursues utility maximization, and (2) the individual needs to pay the information-processing cost to acquire more knowledge about the utilities of the destinations. The objective function of the free utility model of the destination choice system with a single origin  (see Eq.~(\ref{eq7})) is mathematically consistent with the Helmholtz free energy in physics. In other words, this destination choice system is analogous to the isothermal and isochoric thermodynamic system that consists of several subsystems in thermal contact with a large reservoir: the number of travelers is analogous to the number of particles; the first term of the free utility model's objective function is analogous to the thermodynamic system's internal energy including potential energy; the information-processing price is analogous to the temperature of the reservoir; the information entropy is analogous to the entropy of the thermodynamic system; the information-processing cost is analogous to the heat transferred between the thermodynamic system and the reservoir; and the marginal utility is analogous to the chemical potential of the subsystem. However, the essence of these two systems is different: the thermodynamic system follows the minimum free energy principle to make the system reach the equilibrium state in which all subsystems have the same chemical potential \cite{Kittel80}, and the maximization free utility of destination choice system is the result of individuals following the equimarginal principle to make choices to maximize their own utility.

Some previous models, including the free cost model \cite{Tomlin68} and the maximum entropy model \cite{Wilson67}, have not shown a clear derivation of the gravity model from microscopic mechanisms \cite{Bar19}. The free cost model presented by Tomlin and Tomlin derived the gravity model by analogy with the Helmholtz free energy, but it did not explain the social gravity law from the perspective of individual choice behavior. In our opinion, this model is not essentially different from the unconstrained gravity model established by direct analogy with Newton's law of universal gravitation. Similarly, Wilson's maximum entropy model on  the gravity model can provide only the most likely macrostate but cannot describe individual choice behavior. The maximum entropy model also requires the prior constraint of the total cost, which is actually the result of individual choice. Additionally, both of these physical analogy models ignore the interaction among individuals, which is a ubiquitous phenomenon in social systems \cite{Bar11}. In comparison, we provide a concise explanation for the social gravity law from the perspective of individual behavior. It simultaneously reflects two dominant mechanisms that are common in social systems, namely, individual interaction and bounded rationality. This line of explanation brings us to the idea that the social gravity law is a  phenomenon resulting from bounded rational individuals interacting with each other.

The free utility model can explain not only the social gravity law but also the potential function in game theory from the perspective of individual behavior. The free utility model is a stochastic potential game model \cite{Goeree99} in mathematical form. The objective function of the potential game model proposed by Monderer and Shapley was established by analogy with the potential function in physics \cite{Monderer96}. The variation in a player's individual payoff due to changes in the player's strategy is equal to the variation in the potential function \cite{Grauwin09, Lemoy11}. Every equilibrium strategy profile of the potential game maximizes the potential function. Under these circumstances, all players have the same payoff, and they cannot unilaterally change their strategy to increase their respective payoff. However, Monderer and Shapley raised a question about the interpretation of potential function: ``What do the firms try to jointly maximize?" \cite{Monderer96}. In fact, the potential function is not something that players try to jointly maximize. Now we know that players will maximize only their own utility (i.e., the payoff) through the modeling process of the free utility model. A player's optimal choice strategy is to make the marginal utility of each alternative equal. When all players follow this equimarginal principle, the sum of the integral of the marginal utility of each alternative (called total utility in economics) is naturally maximized. In other words, the maximum potential function is the result of players' optimal choice strategy but is not a goal that the players pursue jointly.

The free utility model can also provide an explanation for the objective function of the classic traffic assignment model in transportation science. From Eq.~(\ref{eq13}), we can see that the free utility model for the transportation network is mathematically consistent with the SUE model \cite{Fisk80}. Without considering the information-processing cost, the free utility model is mathematically consistent with the classic user equilibrium (UE) model established by Beckmann \cite{Beckmann55}. The equilibrium solution of the UE model, in which all paths used between each OD pair have equal and minimum costs, is simply the result of  the optimal path choice strategy selected by travelers following the equimarginal principle. The objective function of the UE model, i.e., the sum of the integral of the marginal utility (negative cost) of each link, is actually the negative free utility without information-processing cost. This provides new insights into UE and SUE models.

Finally, although we use individual destination choice behavior in the transportation system as background to explain the social gravity law, such behavior of selecting alternatives exists not only in the transportation system but also in systems where the spatial interaction patterns follow the social gravity law. For example, population migration involves selecting locations as residences, social interaction involves selecting people as friends and scientific collaboration involves selecting researchers as partners. In these systems, individuals tend to select alternatives with relatively high activity and relatively low distance, and changes in the number of individuals who select the same alternative will affect the variation in the utility of the alternative. For example, an increase in commodity competitors for the same product will lead to a decrease in the product price, and an increase in collaborators with one scientist will lead to a decrease in cooperation intensity between this scientist and her collaborators. The spatial interaction behavior in these different systems can be described by the free utility model. Not only that, the free utility model of the destination choice system with a single origin (see Eq.~(\ref{eq7})) can be used for generalized problems of human choice, where homogeneous individuals select from multiple alternatives with utility associated with the choices. Moreover, using the network method, we can extend the free utility model to individuals of different types (i.e., heterogeneous individuals). The network in Fig.~\ref{fig2} can be regarded as a case in which two types of individuals select from among three alternatives. In this network, the utility of the solid line represents the utility component of the alternative for the corresponding type of individual, and the utility of the dashed line represents the utility component of the alternative for both types of individuals. In short, the free utility model that explains the social gravity law not only helps us deeply understand collective choice behavior patterns emerging from the interaction of bounded rational individuals but also shows potential application in predicting, guiding or even controlling human choice behavior in various complex social systems.

\section*{{Acknowledgments}}
This work was supported by the National Natural Science Foundation of China (Grant Nos. 71822102, 71621001, 71671015).

\appendix
\setcounter{section}{1}
\section*{Appendix}
\setcounter{equation}{0}
\label{appx}
\renewcommand{\theequation}{A.\arabic{equation}}
The Lagrangian expression of the optimization problem in Eq. (\ref{eq10})  is

  \begin{equation}
\label{eqdoub1}
\begin{aligned}
\max L(T_{ij}) = &\sum\limits_{i} \sum\limits_{j} \int_0^{T_{ij}} (A_j - c_{ij}- \gamma \ln x) \mathrm{d} x\\& + \tau \sum_i S_i  \\ &+\sum\limits_{i}\lambda_i (\sum\limits_j T_{ij} - O_i) \\& + \sum\limits_{j}\mu_j (\sum\limits_i T_{ij} - D_j),
\end{aligned}
\end{equation}
where $\lambda_i$ and $\mu_j$ are Lagrange multipliers. Since
\begin{equation}
\label{eqdoub2}
\begin{aligned}
S_i &= -O_{i} \sum\limits_j p_{ij}\ln p_{ij}\\
&= -O_{i} \sum\limits_j (T_{ij} / O_i)\ln (T_{ij} / O_i)\\
&=-\sum\limits_j T_{ij}\ln (T_{ij} / O_i) \\
&= -\sum\limits_j T_{ij}\ln T_{ij} + O_i\ln O_i\\
&=-\sum\limits_j T_{ij}\ln T_{ij} + O_i + O_i\ln O_i - O_i\\
&=-\sum\limits_j (T_{ij}\ln T_{ij} - T_{ij}) + O_i\ln O_i - O_i,
\end{aligned}
\end{equation}
the partial derivative of the Lagrangian expression Eq. (\ref{eqdoub1}) with respect to $T_{ij}$ is 
   \begin{equation}
\label{eqdoub3}
\frac{\partial L}{\partial T_{ij}}= A_j - c_{ij}- \gamma \ln T_{ij} - \tau \ln T_{ij} + \lambda_i+ \mu_j = 0,
\end{equation}
therefore,
 \begin{equation}
\label{eqdoub4}
T_{ij}  = \mathrm{e}^{(A_j - c_{ij} + \lambda_i+ \mu_j) / (\gamma + \tau)}.
\end{equation}
Since 
 \begin{equation}
\label{eqdoub5}
\begin{aligned}
O_i =& \sum\limits_j T_{ij} = \sum\limits_j \mathrm{e}^{(A_j - c_{ij} + \lambda_i+ \mu_j) / (\gamma + \tau)}\\
=& \mathrm{e}^{\lambda_i / (\gamma + \tau)} \sum\limits_j \mathrm{e}^{(A_j - c_{ij} + \mu_j) / (\gamma + \tau)}
\end{aligned}
\end{equation}
and 
 \begin{equation}
\label{eqdoub6}
\begin{aligned}
D_j = &\sum\limits_i T_{ij} = \sum\limits_i \mathrm{e}^{(A_j - c_{ij} + \lambda_i+ \mu_j) / (\gamma + \tau)}\\
=& \mathrm{e}^{(A_j + \mu_j) / (\gamma + \tau)} \sum\limits_i \mathrm{e}^{(\lambda_i - c_{ij}) / (\gamma + \tau)},
\end{aligned}
\end{equation}
we obtain
 \begin{equation}
\label{eqdoub7}
\mathrm{e}^{\lambda_i / (\gamma + \tau)} = O_i /  \sum\limits_j \mathrm{e}^{(A_j - c_{ij} + \mu_j) / (\gamma + \tau)}
\end{equation}
and
 \begin{equation}
\label{eqdoub8}
\mathrm{e}^{ (A_j + \mu_j) / (\gamma + \tau)} = D_j /  \sum\limits_i \mathrm{e}^{(\lambda_i - c_{ij}) / (\gamma + \tau)}.
\end{equation}
Let $a_i =\mathrm{e}^{\lambda_i / (\gamma + \tau)} / O_i$ and  $b_j =\mathrm{e}^{ (A_j + \mu_j) / (\gamma + \tau)} / D_j $; 
Eq. (\ref{eqdoub4}) can then be  rewritten as Eq. (\ref{eq11}).

In the actual calculation, $a_i$ and $b_j$ are two sets of interdependent balancing factors, i.e., 
 $a_{i}=1 / \sum_{j} b_{j}D_{j}\mathrm{e}^{-c_{ij} / (\gamma + \tau)}$ and $b_{j}=1 / \sum_{i} a_{i}O_{i}\mathrm{e}^{-c_{ij} / (\gamma + \tau)}$.
Thus, the calculation of one set requires the values of the other set: start with all $b_j = 1$, solve for $a_i=1/\sum_{j}b_{j}D_{j}\mathrm{e}^{-c_{ij} / (\gamma + \tau)}$,
use these values to re-estimate  $b_j=1/\sum_{i}a_{i}O_{i}\mathrm{e}^{-c_{ij} / (\gamma + \tau)}$, and repeat until convergence of the two sets is achieved \cite{Dios11}.

\end{document}